\documentclass[a4paper,10pt]{article}
\usepackage{graphicx}

\title{A note on measurement of network vulnerability under random and intentional attacks}
\date{}

\author{A.Yazdani\thanks{Corresponding author. Email: a.yazdani@cranfield.ac.uk
\vspace{6pt}}, P. Jeffrey\\\vspace{6pt}
{\em{\small School of Applied Sciences, Cranfield University, MK43 0AL, UK.}}}

\begin{document}
\maketitle

In this paper we propose an alternative approach for the assessment of network vulnerability under random and intentional attacks as compared to the results obtained from the \textquotedblleft vulnerability function\textquotedblright given by Criado et al. [Int. J. Comput. Math., 86 (2)(2009), pp. 209-218]. By using spectral and statistical measurements, we assess robustness as the antonym to vulnerability of complex networks and suggest a tentative ranking for vulnerability, based on the interpretation of quantified network characteristics. We conclude that vulnerability function, derived from the network’s degree distribution and its variations only, is not general enough to reflect the lack of robustness due to the specific configurations in graphs with hierarchical or centralized structures. The spectral and statistical metrics, on the other hand, capture different aspects of network topology which provide a more thorough assessment of network vulnerability.\bigskip


\section{Introduction}
Complex networks have several interdependent elements with non-trivial layouts in which the degree of complexity depends on the configurations and the level of interaction between those elements. One important observation giving rise to studies of the topology of complex networks is that \textquotedblleft the structure affects function\textquotedblright [1] and therefore the architecture of the network can be used to understand and predict dynamical processes that affect its performance and flow distribution as well as network stability and tolerance. Researchers investigating the structure and behavior of complex networks have proposed different methodologies and metrics to (i) identify vulnerable and weak points (ii) locate their positions in the networks and (iii) assess network robustness and tolerance to errors and attacks [2-6]. These metrics can be classified as basic connectivity metrics, spectral measurements and statistical measurements. Basic connectivity metrics include vertex- (node-) connectivity and edge- (link-) connectivity [7] that represent network cohesion and adhesion and its sensitivity to the removal of nodes and links. Spectral measurements are those which relate network topology to connectivity strength and graph cohesion through analysis of the spectrum of the network’s adjacency matrix. Examples of well-established spectral metrics include algebraic connectivity [8] to describe network robustness against node and link failures [8-9], and spectral gap analysis to quantify expansibility of the network and the property of \textquotedblleft Good Expansion\textquotedblright [10]. Statistical measurements of complex networks on the other hand provide quantifications for rather intuitive underlying properties through studying the most frequent patterns and building blocks of the networks. Notable examples of this class of measurements include node-degree distribution, clustering coefficient, average path-length, node betweenness and central-point dominance proposed by different authors and listed in [11] along with numerous other metrics.

One recent addition to the stable of vulnerability metrics which relates robustness to network regularity has been proposed by Criado et al. [12] and has been used on several occasions to support network vulnerability assessments [12-14]. This metric, called the \textquotedblleft vulnerability function\textquotedblright, is based on an axiomatic rather than intuitive approach to the concept of network robustness and provides a relatively easy to calculate measure of network vulnerability. In outline, this is a normalized measure which provides values close to one for the most vulnerable graphs decreasing to values closer to zero for more robust graphs. Boundary values of zero or one may only be attained asymptotically and consequently there exist graphs which have vulnerability values arbitrarily close to either zero or one. This may not necessarily happen in graphs with a fixed size, requiring the asymptotic computation to take place in polynomial time [12]. The rationale behind the definition is that any vulnerability measure should be invariant under isomorphisms and that it should not increase by adding only edges (links) to the original graph (although this is not necessary when vertices are added). The definition of the vulnerability function $\nu_\sigma:%
\Omega
\rightarrow [0,1]$ for the set $%
\Omega
$ of all possible finite graphs is given in Criado et al. [12] as a function verifying the following properties:\bigskip
\\
$(i)$ $\nu_\sigma$ is invariant under isomorphisms.
\\
$(ii)$ $\nu_\sigma(G^{^{\prime }})\geq \nu_\sigma(G)$ if $G$ is
obtained from $G^{^{\prime }}$ by adding edges.
\\
$(iii)$ $\nu_\sigma(G)$ is computable in polynomial time with respect to the number of vertices of $G$.\bigskip
\\\
Based on these criteria the authors [12] have proposed the vulnerability of a graph $G$ as:
\begin{equation}
\nu_\sigma(G)=\exp (\frac{\sigma }{n}+n-\left\vert E\right\vert -2+\frac{%
2}{n})  \label{Equation 1}
\end{equation}%
where n is the number of vertices, $\left\vert E\right\vert $ is the number of edges and $\sigma $\ is the standard deviation of the degree distribution. The measure has subsequently been deployed as an accurate and computable definition of network vulnerability [13-14]. 

However, we argue that the vulnerability measure described above constitutes only a partially reliable comparison for different types of network. Specifically, whilst our experiments show that the definition is supported by some examples that reasonably compare the vulnerability of networks possessing more or less the same size and structures, it fails to properly rank vulnerability when networks of different sizes and structures are compared. In particular, and as will be demonstrated, complete regular graphs of relatively low order are found to be more vulnerable by this measure than graphs which are intuitively (and technically) understood as more vulnerable due to the existence of bridges and cut vertices whose removal disconnects the network causing isolation of some nodes from the source(s). Notably, intuitive concepts of graph strength such as vertex connectivity (cohesion) and edge connectivity (adhesion) are not captured by this measure. Therefore, it is observed that, taking robustness as the antonym for vulnerability to random failures or intentional attacks, graph cohesion as a fundamental constituent of robustness is ignored by this vulnerability function which yields computations that are incompatible with other robustness measurements reported in the literature. It is also reported by Holmgren [15] that the vulnerability function fails to capture actions taken to enhance the resilience of networks. These observations lead us to an alternative approach based on the quantification of network invariants and statistical properties followed by heuristic interpretations. In the following sections, we provide two examples of four network models with different layouts and structures, derive the specific set of relevant spectral and statistical network metrics that we use to relate network architecture to robustness, and utilize these metrics to assess network vulnerability against scenarios of node and link failures. We then compare our findings with the outcome of applying the vulnerability function in Equation (1) and comment on the function’s usefulness. Furthermore, we make recommendations on the interpretation and utilization of spectral and statistical measurements as a basis for the structural analysis of networks so as to obtain a comprehensive and more generically applicable assessment of robustness and vulnerability. 

\section{Measures of network robustness and vulnerability}
Let $G=(N,E)$ be a connected unweighted network consisting of the set of vertices $N$ and the set of edges $E$, with distinct vertices $S$ for the source and $T$ for the sink(s). Following [7], vertex-connectivity $\kappa(G)$ also known as graph cohesion is defined as the smallest number of vertices whose removal leaves the graph disconnected. Edge-connectivity $\mu(G)$  also known as graph adhesion is similarly defined to be the smallest number of edges whose removal disconnects the graph. Algebraic connectivity $\lambda_2 (G)$ first introduced in [8] and studied in [9] for its scope of applicability, is the second smallest eigenvalue of the Laplacian matrix of a network. The Laplacian matrix of $G$ with n nodes is a $n$$\times$$n$ matrix $Q$=$\Delta$-$A$, where $\Delta$= diag($d_i$) and $d_i$ is the degree of node $i$ and $A$ is the adjacency matrix of $G$. The algebraic connectivity $\lambda_2 (G)$ is a positive value whose magnitude indicates the level of connectivity in the graph. Larger values of algebraic connectivity represent higher graph robustness against efforts to decouple parts of the graph. Spectral Gap $s(G)$ is defined as the difference between the first and second eigenvalues of the adjacency matrix A with sufficiently large values of spectral gap being a necessary condition for Good Expansion properties. Low values of Spectral Gap indicate a lack of Good Expansion properties usually represented by bridges, cut vertices and network bottlenecks [10]. 

The statistical measurements are widely used in the complex networks literature to quantify network structure in relation with vulnerability. Here, we employ some of these statistical measurements to provide support for our analysis of network vulnerability based on the spectral measurements. These include; link density $q(G)$ defined as the fraction between the total and the maximum number of possible links [9] to assess availability of alternative routes in the case of failure, mean node-degree $\prec$$k$$\succ$ as a measure of average global connectivity [11] or nodal strength, average path-length $l(G)$ [11] reflecting the ease of access in the network and reciprocally related to the efficiency in dispatching flow, and central-point dominance $C_B (G)$ defined as the average betweenness difference of the most central point and all others [16] which quantifies the extent to which a network is decentralized or concentrated (around a centre). A detailed overview of the applicability and relevance of these metrics to topology-related network vulnerability can be found in [11] and references therein.

\section{Examples on network robustness analysis}
\subsection{Example 1}
Consider the complete graph of order four $K_4$ $(n=4,|E|=6)$ and the hierarchically structured graph $G_1$ $(n=8,|E|=10)$ illustrated below (Figure 1). Without loss of generality, it can be assumed that these graphs represent flow networks with a single source $S$ and a single sink $T$ and with the same weight and capacity allocated to every node and link. It is readily observed from Equation (1) that $\nu_\sigma$$(K_4)$=0.0302 and $\nu_\sigma$$(G_1)$=0.0269 which suggests that $K_4$  is more vulnerable than $G_1$ with respect to random and intentional attacks. However, we show that this is not a reliable finding by applying the following attack scenario to demonstrate that $G_1$ has smaller edge and vertex connectivity values than $K_4$. While removal of the only edge attached to the source will render $G_1$ disconnected, no single edge removal will result in disconnection of $K_4$. The higher cohesion and hence greater strength of $K_4$ becomes apparent by considering that, due to the equal impact of the attacks to the edges with the same capacity, a complete loss of access from the source to the sink in $K_4$ requires at least three times more such efforts as compared to $G_1$. A similar disconnection strategy resulting in the removal of the node (and its edges) incident to the source in $G_1$ and comparing that to the total number of required node removals to disconnect $K_4$ demonstrates higher adhesion and greater robustness of $K_4$ which, as far as the robustness and tolerance to failures is concerned, contradicts the vulnerability measurements computed by Equation (1).

\begin{figure}[h]
\centering
\includegraphics[width=11cm]{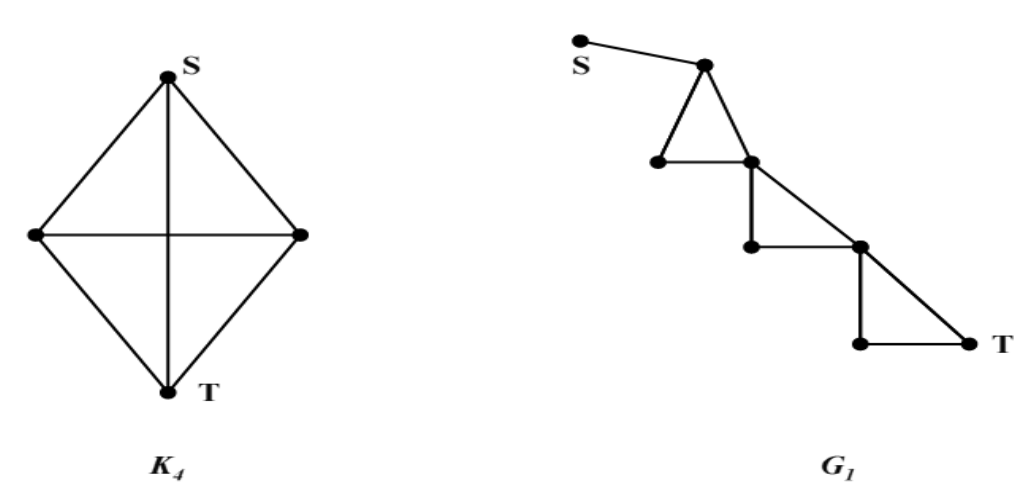}
\caption{Comparison of the structures for $K_{4}$ and $G_{1}$}
\label{Figure 1}    
\end{figure}

\subsection{Example 2}

The next example provides a comparison of the vulnerability between the complete graph $K_5$ $(n=5,|E|=10)$ of order five and $G_2$  $(n=13,|E|=21)$ a graph with loosely connected community structures (Figure 2). Similar flow network assumptions are made in this example with the additional assumption of multiple sinks in each network. Equation (1) yields relative network vulnerability values of $\nu_\sigmaσ$$(K_5)$=0.0014 and $\nu_\sigmaσ$$(G_2)$=0.0001 which suggests that $K_5$ is more vulnerable than $G_2$ with respect to random and intentional attacks. However, by employing a strategy similar to the one developed in the previous example, it is found that $K_5$ enjoys a greater degree of cohesion (adhesion) than $G_2$ in light of the fact that disconnection of $G_2$ can take place by only one effort which is the removal of any edge incident to the source while it takes four such efforts to disconnect $K_5$. In other words the existence of bridges and cut vertices (articulation points) in $G_2$ makes this network less tolerant to attacks than $K_5$ which benefits from a complete regular structure with redundant routes from the source to the sinks.

\begin{figure}[h]
\centering
\includegraphics[width=12.5cm]{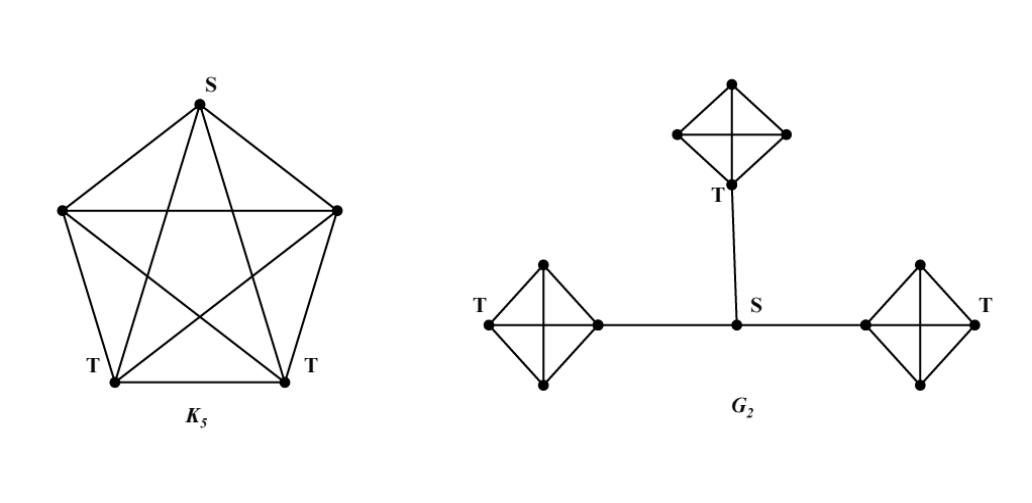}
\caption{Comparison of the structures for $K_{5}$ and $G_{2}$}
\label{Figure 2}       
\end{figure}

\section{Connectivity and statistical measures compared to vulnerability function}
Numerical calculations presented in Table (1) illustrate the connectivity, spectral and statistical measurements of the studied examples along with vulnerability values calculated for these networks. Being able to directly compare any two networks and rank them according to their level of vulnerability largely depends on the nature of the problem and existence of conclusive measurements followed by heuristic interpretations. In other words a unique graph metric to satisfy all aspects of connectivity and robustness may not exist, as also reported elsewhere [17], and therefore the appropriate metric(s) for making a judgment on network robustness need to be selected from a set of measurements that capture different appropriate information. However, derived measurements for particular examples given in the previous section do support a tentative ranking for network vulnerability. 

\begin{table}[ht]
\caption{\scriptsize Comparison of the networks based on vulnerability, connectivity and statistical measurements}
\centering
\small
\begin{tabular}{p{0.8cm}p{1cm}p{1cm}p{1cm}p{0.9cm}p{0.9cm}p{0.9cm}p{0.9cm}p{0.9cm}}
\hline
$G$ & $\nu_\sigma$$(G)$ & $s(G)$ & $\lambda_2 (G)$ & $\kappa(G)$ & $q(G)$ & $\prec$$k$$\succ$ & $l(G)$ & $C_B (G)$ \\
\hline
$G_1$  &  2.69  &  0.99  &  0.4    &   1   &  0.36   &  2.5      &  2.00   & 0.46  \\
$G_2$  &  0.01  &  0.27  &  0.21  &   1   &  0.27   &  3.25    &  2.65   & 0.63   \\
$K_4$  &  3.02  &  2       &  4      &   3   &  1        &  3	      &  1.00   & 0         \\
$K_5$  &  0.14  &  3       &  5      &   4   &  1        &  4 	      &  1.00   & 0         \\ 
\hline
\end{tabular}
\label{table: Table (1)}
\end{table}

As shown in Table (1), the vulnerability function in Equation (1) ranks the networks in the order of $\nu_\sigma$$(G_2)$$\leq$$\nu_\sigma$$(K_5)$$\leq$$\nu_\sigma$$(G_1)$$\leq$$\nu_\sigma$$(K_4)$ with network $G_2$ the least vulnerable and $K_4$ the most vulnerable. However, based on the other metrics, it can be strongly suggested that $G_2$ is the most vulnerable with $G_1$, $K_4$ and $K_5$ being the less vulnerable networks. 

\section{Discussion}
It is observed that the trend in the values of algebraic connectivity is perfectly followed by all other connectivity and statistical metrics with only the exception of the Mean node-degree. An interpretation of this observation is that graph robustness, as reflected by vertex connectivity and algebraic connectivity, is replicated by Good Expansion properties quantified by larger spectral gaps. The existence of alternative and redundant routes (represented by link density) along with smaller average distances, increases the networks’ survivability and facilitates reachability in the networks that results in the improvement of robustness to failures. Central-point dominance provides some insight to network design and tolerance to failures by measuring the extent to which one individual node controls the flow in the network. Higher central-point dominance, as seen in the case of $G_2$, may lead to higher vulnerability due to the consequences of failing central point(s), contrary to more decentralized (e.g. $G_1$) or completely distributed networks (e.g. $K_4$ and $K_5$). The trend in the Mean node-degree is in an agreement with other metrics for the most invulnerable graphs $K_4$ and $K_5$ while the differences observed otherwise seem to be more related to the size and degree distribution than the graph layouts.  It is worth noting that vertex and edge connectivity are configuration-only measures independent of the number of graph vertices. Algebraic connectivity, on the other hand depends on both the number of nodes and their respective configurations. Given that the vulnerability function in Equation (1) depends on the network degree distribution and its variations [14], a degree of correlation between this measurement and the algebraic connectivity values would be expected theoretically. Its non-existence in this instance remains unexplained.
We also note that the edge connectivity of a subgraph can exceed the overall edge connectivity of the graph. This may happen in networks with hierarchical structures (e.g. $G_1$) or in hub-spoke networks with loosely related clusters (e.g. $G_2$). Moreover, it is observed that mesh-like distributed networks enjoy better accessibility among the nodes and therefore have an overall larger degree of reliability and lower vulnerability than centralized and hub-spoke networks. This is due to the fact that mesh-like structures posses a certain level of path-redundancy and they resemble the most invulnerable graphs i.e. complete regular graphs, at least locally. Hierarchical and hub-spoke structures on the other hand, fall into the category of sparse Non-Good Expansion networks which typically have low connectivity values along with the presence of the bottlenecks and cut vertices. These are networks with relatively small spectral gaps resulting from the small difference between the first and the second eigenvalues of the network adjacency matrix. Dekker [18] suggests that the most robust networks are those optimally connected in the sense of having vertex and edge connectivity as high and average distance as low as possible, with symmetric and node-similar structure and balanced distribution of the flow along the links. These properties can be perfectly found in complete graphs $K_4$ and $K_5$ making them the most robust networks among the others, as discussed. The vulnerability of the two other examples can be assessed based on the other criteria listed in this work or by considering other network metrics and characteristics. These observations suggest that any quantification of vulnerability needs to adequately reflect different network patterns and structures as well as the variations in graph size, vertex degree and nodal configurations. 

\section{Conclusions}

Examples presented in this paper showed that the discussed “vulnerability function” proposed in [12] does not provide a reliable measurement of network vulnerability and in particular does not capture important aspects of structural vulnerability such as connectivity and cohesion. Therefore, construction of a closed-form mathematical expression to quantify network vulnerability will very much depend on the successful incorporation of different network measurements including those discussed in this work. As demonstrated, an alternative approach based on the derivation of spectral and statistical measurements of networks can be developed by which different aspects of network topology in relation to robustness and vulnerability may be investigated.  However, the extent to which these metrics explain non-topological aspects of network reliability and redundancy needs to be scrutinized. Overall, a thorough and realistic assessment of network vulnerability will depend on the nature and objectives of the analysis as well as the existence of conclusive measurements followed by heuristic interpretations.

\section*{Acknowledgements}
This work was conducted with the support of a Leverhulme Trust grant on applications of graph theory and complex networks to understanding robustness in water distribution systems.


\begin{thebibliography}{18}


\bibitem[1]{1} S. H. Strogatz, Exploring complex networks, Nature, 410 (2001), pp. 268-276.

\bibitem[2]{2} V. Gol’dshtein, G. A. Koganov, G. I. Surdutovich, Vulnerability and Hierarchy of Complex Networks, Cond. Mat., (2004), Art. No. 0409298.

\bibitem[3]{3} R. Albert, H. Jeong, A. L. Barabasi, Error and Attack Tolerance of Complex Networks, Nature, 406 (2000), pp. 378-382.

\bibitem[4]{4} V. Latora, V., M. Marchiori, Vulnerability and Protection of Critical Infrastructure, Cond. Mat., (2004), Art. No.  0407491.

\bibitem[5]{5} G. Paul, T. Tanizawa, S. Havlin, H. E. Stanley, Optimization of robustness of complex networks, Eur. Phys. J. B, 38 (2004), pp. 187-191.

\bibitem[6]{6} E. Bompard, R. Napoli, F. Xue, Analysis of structural vulnerabilities in power transmission grids, International Journal of Critical Infrastructure Protection, 2 (2009), pp. 5-12.

\bibitem[7]{7} G. Chartrand, L. Lesniak, Graphs \& Digraphs, Third Edition, Chapman \& Hall, 1996. 

\bibitem[8]{8} M. Fiedler, Algebraic connectivity of graphs, Czechoslovak Mathematical Journal, 23 (1973), pp. 298-305. 

\bibitem[9]{9} A. Jamakovic, S. Uhlig, On the relationship between the algebraic connectivity and graph’s robustness to node and link failures, Proceedings of the 3rd EURO-NGI Conference on Next Generation Internet Network, Trondheim, Norway, (2007), pp. 96-102.

\bibitem[10]{10} E. Estrada, Network robustness to targeted attacks. The interplay of expansibility and degree distribution, Eur. Phys. J. B., 52 (2006), pp. 563-574.

\bibitem[11]{11} L. F. Costa, F. A. Rodrigues, G. Travieso, P. R. Boas, Characterization of complex networks: A survey of measurements, Advances in Physics, 56 (1) (2007), pp. 167-175.

\bibitem[12]{12} R. Criado, J. Flores, B. Hernandez-Bermejo, J. Pello, M. Romance, Effective Measurement of Network Vulnerability Under Random and Intentional Attacks, J. Math. Model. Algorithms, 4 (2005), pp. 307-316.  

\bibitem[13]{13} R. Criado, J. Flores, M. I. Gonzalez-Vasco, J. Pello, Choosing a leader on a complex network, J. Comput. Appl. Math. 204 (2007), pp. 10-17.

\bibitem[14]{14} R. Criado, J. Pello, M. Romance, M. Vela-Perez, Improvements in performance and security for complex networks, Int. J. Comput. Math., 86 (2) (2009), pp. 209-218.

\bibitem[15]{15} A. J. Holmgren, Using Graph Models to Analyze the Vulnerability of Electric Power Networks, Risk Analysis, 26 (4) (2006), pp. 955-969.

\bibitem[16]{16} L. C. Freeman, A Set of Measures of Centrality Based on Betweenness, Sociometry, 40 (1) (1977), pp. 35-41. 

\bibitem[17]{17} A. Bigdeli, A. Tizghadam, A. Leon-Garcia, Comparison of Network Criticality, Algebraic Connectivity, and other Graph Metrics, SIMPLEX09, 1st Annual Workshop on Simplifying Complex Network for Practitioners, Co-located with NetSci'09, (2009), Art. No. 4.

\bibitem[18]{18} A. H. Dekker, Simulating Network Robustness for Critical Infrastructure Networks, Proceedings of 28th Australasian Computer Science Conference, Newcastle, Australia, January (2005).

\end{thebibliography}
\end{document}